\documentclass{aa}  
\usepackage{graphicx, bm}
\usepackage{txfonts, subcaption, hyperref, float, amsmath, amssymb,calrsfs, stackengine,scalerel}
\DeclareMathAlphabet{\pazocal}{OMS}{zplm}{m}{n}

\usepackage{hyperref, xcolor}
\hypersetup{
   colorlinks=true,
   linkcolor=blue,
   filecolor=blue,      
   urlcolor=blue,
   citecolor=blue
}

\begin{document}

   \title{An integration-free method for calculating curl and divergence in space plasmas using multi-spacecraft data}
   \titlerunning{An integration-free method for calculating curl and divergence in space plasmas}

   \subtitle{}

   \author{Rohan Singh, Supratik Banerjee 
          \and Arijit Halder
          }

    \authorrunning{Singh, R., et al.}

   \institute{Department of Physics, Indian Institute of Technology, Kanpur, 208016, India \\
              \href{mailto:rohansingh23@iitk.ac.in}{rohansingh23@iitk.ac.in},  \href{mailto:sbanerjee@iitk.ac.in}{sbanerjee@iitk.ac.in}, \href{mailto:harijit@iitk.ac.in}{harijit@iitk.ac.in}
             }

   \date{Received Month XX, XXXX; accepted Month XX, XXXX}

 
  \abstract
  { The knowledge of local spatial gradients (curl, divergence etc.) is crucial to examine the three-dimensional variation of flow fields including velocity and magnetic fields in space plasmas like the solar wind. Here we propose a simple method to calculate the same using the \textit{in-situ} data of multi-spacecraft systems. Unlike the popular \textit{Curlometer} method which depends on the vector integration theorems, our integration-free method is based on the construction of a local orthonormal coordinate system and the associated finite difference approximations. The \textit{Curlometer} is applicable to a four spacecraft system arranged in a tetrahedron and yields a single volume-averaged estimate of the curl. Using our proposed method over 107 intervals of MMS (NASA) data, on the other hand, we successfully calculate the spatial derivatives at the position of each spacecraft of the tetrahedron and a three-spacecraft (non-collinear) subset of the same. The average value of all the curls calculated for a given tetrahedron shows an excellent agreement (correlation coefficient $\sim 0.99$) with the curls calculated using \textit{Curlometer} formula. The quality of the calculated curl (using our method) is found to improve if the   spacecraft configuration approaches a regular tetrahedron. The current framework facilitates investigation of turbulent heating rates and the exploration of local flow features like Beltramization, existence of current sheets, etc., in present and future multi-spacecraft mission, including those involving more than four spacecraft.  }


   \maketitle
%

\section{Introduction} \label{sec1}

The signature of turbulence in the solar wind is clearly revealed through the fluctuations in its electromagnetic fields and plasma variables (density, velocity \textit{etc.}), extending over a wide range of spatial and temporal scales. Turbulence is believed to play a crucial role in explaining the acceleration and the local heating of the solar wind \citep{Verma_1995, Matthaeus_1999, Sorriso-Valvo_2007, Vasquez_2007, Carbone_2009, Banerjee_2016, Chen_2021}. For length scales larger than the ion inertial scale, several features of solar wind turbulence are adequately explained by the framework of magnetohydrodynamic (MHD) turbulence \citep{Grappin_1990, Goldstein_1995, Horbury_2005, Bruno_2013}. 

\begin{figure}
    \centering
    \includegraphics[width=0.5\linewidth]{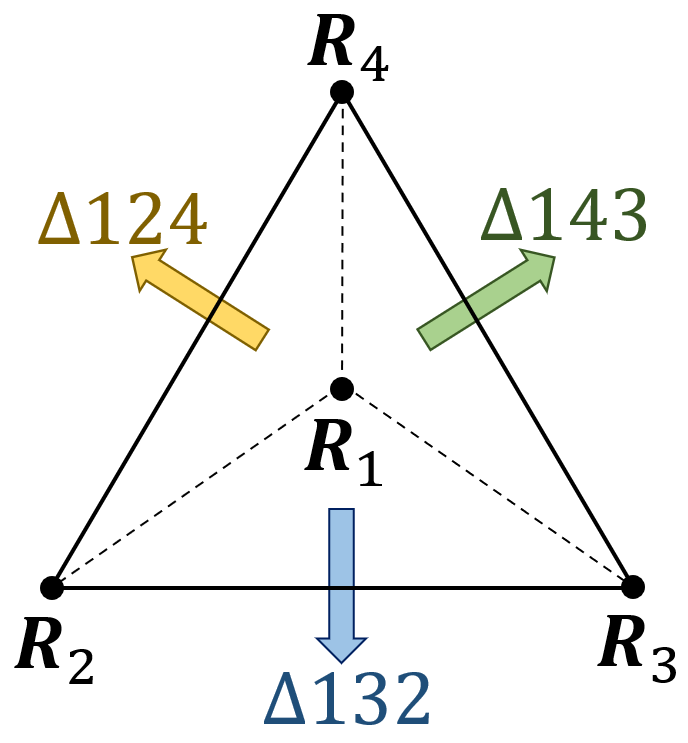}
    \caption{ Schematic diagram of a tetrahedral multi-spacecraft system with the spacecraft being located at the vertices (with positions denoted by $\bm{R}$). The thick arrows denote the direction of the surfaces common to MMS 1 (located at $\bm{R}_1$). }
    \label{curlometer_tetrahedron}
\end{figure}
An exact measure of the turbulent heating rate can be obtained by the exact relation of incompressible MHD turbulence given by $\bm{\nabla_r}\cdot\pazocal{F} = - 4 \varepsilon$ \citep{Politano_1998}, where $\varepsilon$ is the energy cascade rate (equal to the turbulent heating rate) and 
\begin{equation}
    {\pazocal{F}} = \left\langle \delta \bm{u}\ \left [(\delta \bm{u})^2 + (\delta \bm{b})^2\right ] - 2 \delta \bm{b}\ (\delta \bm{u}\cdot \delta \bm{b}) \right\rangle,
\end{equation}
where $\bm{u}$ is the fluid velocity and $\bm{b} = \bm{B}/\sqrt{\mu_0}$ with $\bm B$ being the magnetic field, $\mu_0$ the free space permeability, and $\delta$ represents the spatial increment between two points separated by $\bm r$.  In case of isotropic turbulence, the above exact relation can be integrated over a sphere of radius $r$ to obtain $\pazocal{F}_r = -(4/3)\varepsilon r$. But, the non-negligible mean magnetic field of the solar wind incorporates anisotropy in the energy transfer \citep{Matthaeus_1996, Horbury_2008, Podesta_2009, Telloni_2019, Deepali_2021, Sakshee_2022}. An alternative exact relation, which comes out to be particularly useful to calculate the turbulent heating rate without making the assumption of isotropy, is given by \citep{Banerjee_2017} 
\begin{equation}
\left\langle \delta(\bm{u} \times \bm{\omega})\cdot\delta\bm{u} + \delta(\bm{j} \times \bm{b})\cdot\delta\bm{u} + \delta(\bm{u} \times \bm{b})\cdot\delta\bm{j} \right\rangle = 2\varepsilon,
\end{equation}
where $\bm{\omega} = \bm{\nabla}\times\bm{u}$ and $\bm{j} = \bm{\nabla}\times\bm{b}$. In addition to $\bm u$ and $\bm b$, one therefore needs to calculate $\bm \omega$ and $\bm j$ in order to measure $\varepsilon$. The calculation of the curl is not possible using the data from a single-spacecraft but requires a multi-spacecraft system. Various methods to calculate curls using multi-spacecraft mission have been proposed in the past, including the \textit{Curlometer} technique \citep{Dunlop_1988}, the linear interpolation method \citep{Chanteur_1998}, the least squares method \citep{Harvey_1998}, the geometrical method \citep{Shen_2023} etc. All the methods are based on integral calculus and the \textit{Curlometer} \citep{Dunlop_1988} technique has become the most popular of all the aforesaid methods. 
According to this technique, for a given tetrahedral multi-spacecraft structure (Cluster and MMS), as shown in Fig.~ \ref{curlometer_tetrahedron}, we first take the triangular surface $\Delta 132$ and use 
\begin{align}
&{\boldsymbol \nabla} \times {\bm B} = \mu_0 {\bm J} \implies \oint {\bm B} \cdot d {\bm \ell}= \mu_0 \iint\limits_{\Delta 132} \bm{J\cdot dS} \nonumber\\
&\implies \int\limits_1^3 \bm{B\cdot} d {\bm \ell} + \int\limits_3^2 \bm{B\cdot} d {\bm \ell} + \int\limits_2^1 \bm{B\cdot} d {\bm \ell} = \mu_0 \iint\limits_{\Delta 132} \bm{J\cdot dS},
\end{align}
where ${\bm J}$ is the current density. In the next step, for every line integral, a uniform magnetic field, with strength being equal to the average of the magnetic field strengths at the two extremities, is assumed. In addition, if we assume an average current ${\bm J}$ for the entire tetrahedron, the above expression can be approximately expressed as 
\begin{align}
    &\left ( \frac{\bm{B}_1 +\bm {B}_3}{2} \right )\cdot (\bm{R}_3 - \bm{R}_1) + \left ( \frac{\bm{B}_3 + \bm{B}_2}{2} \right )\cdot (\bm{R}_2 - \bm{R}_3) \nonumber \\ &+ \left ( \frac{\bm{B}_2 + \bm{B}_1}{2} \right )\cdot (\bm{R}_1 - \bm{R}_2)  = \mu_0 \bm{J} \cdot \left ( \frac{\bm{R}_{31} \times \bm{R}_{21}}{2} \right ) \nonumber \\
    & \implies \bm{B}_{31}\cdot  \bm{R}_{21} - \bm{B}_{21} \cdot \bm{R}_{31} = \mu_0 \bm{J}\cdot (\bm{R}_{31} \times \bm{R}_{21}),
\label{J_ABC}
\end{align}
where $\bm{R}_i$ is the position vector of $i^{th}$ spacecraft, $\bm{B}_i$ is the corresponding magnetic field, and for any vector ${\bm A}$, $\bm{A_}{ji}= \bm{A}_j - \bm{A}_i$. 
Following the similar methodology for the surfaces $\Delta124$ and $\Delta143$, we get two other equations similar to the Equation \eqref{J_ABC}, solving which, one finally obtains $\bm{J}$ $i.e.$ the curl of ${\bm B}$ \citep{Dunlop_1988}. 
Evidently, by construction, \textit{Curlometer} technique works only for a four-spacecraft mission in tetrahedral configuration and hence the technique cannot be used if the number of spacecraft is less than four. Furthermore, it provides an averaged estimate of the curl (here ${\bm J}$) over the whole tetrahedron instead of calculating the value of the curl at a particular spacecraft location. 

In this paper, we propose a simple and robust tool that complements existing techniques for calculating the spatial derivatives in space plasmas. The method is based on construction of an orthonormal coordinate system fixed to the spacecraft and does not involve any theorem of vector integration. The proposed Integration-Free Method (IFM hereinafter) does not mandate a tetrahedron-shaped multi-spacecraft system (Cluster and MMS), and as we show, the method is successfully applied to a three spacecraft subset of MMS as well. The formalism is based on finding three orthogonal directions followed by using the basic definition of curl which does not involve any integral theorems of vector calculus.  As a result, we can calculate the curl at specific points (locations of the reference spacecraft) rather than an averaged value for the entire set. This further paves the way for the calculation of higher derivatives of the calculated curl quantities, which is not possible with \textit{Curlometer}.  Finally, we show that the proposed method can also be used to calculate the local divergence of the relevant vector fields, making it a complete toolbox to probe the solar wind and similar systems by means of different sort of spatial derivatives of the field variables.

This paper is organized as follows: Section \ref{sec2} describes the selection of data and the interval details used in this study. The working principle of IFM for the calculation of curl and divergence can be found in Section \ref{sec3}. We then present the result in Section \ref{sec4} for four spacecraft system along with the application to three spacecraft system (suppressing the fourth one). In the same section, we present a comparison between IFM and the \textit{Curlometer} method in terms of the correlation coefficients and the histograms over all the datasets. The section ends with a quantitative study relating the accuracy of the calculated curl to the quality of the tetrahedron. In Section \ref{sec5}, we conclude by summarizing the findings of our study and looking at its future aspects.

\section{Data Selection} \label{sec2}

\begin{figure}[H]
    \centering
    \includegraphics[width= 0.7\linewidth]{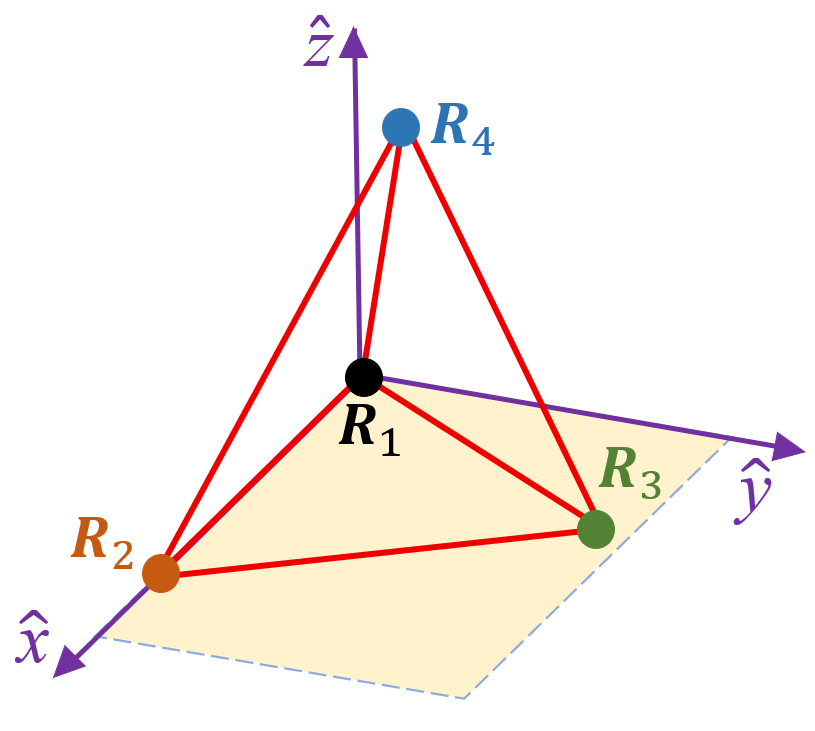}
    \caption{One possible orthonormal coordinate system with MMS 1 at centre and $\hat{x} = \hat{\bm{R}}_{21}$, $\hat{z} =  (\mathbf{R}_{21} \times \mathbf{R}_{31})/\left |\mathbf{R}_{21} \times \mathbf{R}_{31}\right |$, and $\hat{y} = \hat{z}\,\times\,\hat{x}$.}
    \label{first}
\end{figure}

This study has been carried out using the \textit{in-situ} data recorded by MMS (NASA) mission built up of four spacecraft oriented in a tetrahedral formation, each placed at the vertices of the tetrahedron (see Fig. \ref{first}). Onboard MMS, the Fluxgate magnetometers (FGM), consisting of an Analog Fluxgate Magnetometer (AFG) and a Digital Fluxgate Magnetometer (DFG) together, is used to record the magnetic field data at each spacecraft position \citep{Russell_2016, Torbert_2016}. For this study, we use the magnetic field data in the Survey mode with a cadence ranging from 0.0625 to 0.125 seconds . The solar wind velocity data (ions and electrons) is recorded by onboard Fast Plasma Investigation (FPI) Electrostatic Analyser consisting of four dual ion and electron spectrometers (DIS and DES, respectively) per spacecraft \citep{Pollock_2016}. Here, we use the Fast mode data at a cadence of 4.5 seconds. The spacecraft position, magnetic field and ion velocity data are obtained in GSE (Geocentric Solar Ecliptic) coordinate system $\{\hat{i}, \hat{j}, \hat{k}\}$, where $\hat{i}$ is the radial direction pointing towards the Sun from Earth , $\hat{j}$ is the tangential direction opposite to the orbital motion of the Earth in the ecliptic plane and $\hat{k}$ is the direction perpendicular to the ecliptic plane and completes the right-handed orthonormal basis.

In this paper, applying IFM, we calculate the curl for a total of 107 intervals with moderate solar activity between 2015 and 2017. Each dataset is 1-hour long with a resolution of 4.5 seconds (high cadence FGM data is degraded to the resolution of FPI data). These datasets have been selected taking care of the stationarity and avoiding sudden high fluctuations in the data. Any sharp irregular peak in power spectra is filtered out to ascertain regular data. Small data gaps (less than 1 percent) are handled using linear interpolation.

\section{Proposed Method for Curl Calculation} \label{sec3}

Let us calculate the curl of a vector at ${\bm R}_1$ position. The first step is to constitute a new coordinate system with the orthonormal basis $\{\hat{x}, \hat{y}, \hat{z}\}$ fixed to the tetrahedral formation (see Fig. \ref{first}) keeping the origin at MMS 1. One direction can be chosen to be along the line joining MMS 1 and one of the other three spacecraft. The second direction can be defined along the perpendicular to the triangular surface of the tetrahedron, having an edge along the first direction. The third direction is taken to be the one that completes the right-handed coordinate system. For example, we choose the $\hat{x}$ direction to be along $\mathbf{R}_{21}$ and $\hat{z}$ along the normal to the surface $\Delta 123$. The unit vectors are therefore given by $\hat{x} = \mathbf{R}_{21}/|\mathbf{R}_{21}|$, $\hat{z} =  (\mathbf{R}_{21} \times \mathbf{R}_{31})/\left |\mathbf{R}_{21} \times \mathbf{R}_{31}\right |$, and $\hat{y} = \hat{z}\,\times\,\hat{x}$. The new basis vectors, expressed in terms of the GSE coordinates, can be put in the matrix form as, $X = T K$, where\\
$ X = \begin{bmatrix}
        \hat{x}\\
        \hat{y}\\
        \hat{z}
    \end{bmatrix}$, $K= \begin{bmatrix}
        \hat{i}\\
        \hat{j}\\
        \hat{k}
    \end{bmatrix}$, and 
 $T = \begin{bmatrix}
        x_i & x_j & x_k\\
        y_i & y_j & y_k\\
        z_i & z_j & z_k
    \end{bmatrix}$ is the transformation matrix.
If ${\bm A}_n$ represents the vector field $\bm{A}$ at ${\bm R}_n$, the components of the vector in the new coordinates are related to those in the GSE coordinates as  
    \begin{equation}\begin{bmatrix}
        {A_n}_x\\
        {A_n}_y\\
        {A_n}_z
    \end{bmatrix} = \begin{bmatrix}
        x_i & x_j & x_k\\
        y_i & y_j & y_k\\
        z_i & z_j & z_k
    \end{bmatrix}\begin{bmatrix}
        {A_n}_i\\
        {A_n}_j\\
        {A_n}_k
    \end{bmatrix}\end{equation}
Therefore, in $\{\hat{x}, \hat{y}, \hat{z}\}$, the curl of $\bm{A}$ at ${\bm R}_1$ is written as
\begin{equation}
\bm{\nabla} \times \bm{A} {\big |}_1 =
\left(\dfrac{\partial A_z}{\partial y} - \dfrac{\partial A_y}{\partial z} \right) \hat{x} + \left(\dfrac{\partial A_x}{\partial z} - \dfrac{\partial A_z}{\partial x} \right) \hat{y} + \left(\dfrac{\partial A_y}{\partial x} - \dfrac{\partial A_x}{\partial y} \right) \hat{z} \label{eqn6}
\end{equation}

\noindent where 

\begin{subequations}\label{eq:set}
\noindent
\begin{minipage}[t]{0.4\linewidth}
\begin{equation}
  \dfrac{\partial A_z}{\partial x} = \dfrac{{A_2}_z - {A_1}_z}{\mathbf{R}_{21} \cdot \hat{x}},  \label{eq:seta}
\end{equation}
\end{minipage}\hfill
\begin{minipage}[t]{0.4\linewidth}
\begin{equation}
  \dfrac{\partial A_y}{\partial x} = \dfrac{{A_2}_y - {A_1}_y}{\mathbf{R}_{21} \cdot \hat{x}},  \label{eq:setb}
\end{equation}
\end{minipage}

\noindent
\begin{minipage}[t]{0.4\linewidth}
\begin{equation}
  \dfrac{\partial A_z}{\partial y} = \dfrac{{A_3}_z - {A_1}_z}{\mathbf{R}_{31} \cdot \hat{y}}, \label{eq:setc}
\end{equation}
\end{minipage}\hfill
\begin{minipage}[t]{0.4\linewidth}
\begin{equation}
  \dfrac{\partial A_x}{\partial y} = \dfrac{{A_3}_x - {A_1}_x}{\mathbf{R}_{31} \cdot \hat{y}},  \label{eq:setd}
\end{equation}
\end{minipage}

\noindent
\begin{minipage}[t]{0.4\linewidth}
\begin{equation}
 \dfrac{\partial A_y}{\partial z} = \dfrac{{A_4}_y - {A_1}_y}{\mathbf{R}_{41} \cdot \hat{z}},  \label{eq:sete}
\end{equation}
\end{minipage}\hfill
\begin{minipage}[t]{0.4\linewidth}
\begin{equation}
  \dfrac{\partial A_x}{\partial z} = \dfrac{{A_4}_x - {A_1}_x}{\mathbf{R}_{41} \cdot \hat{z}}. \label{eq:setf}
\end{equation}
\end{minipage}
\end{subequations}
\vspace{0.2cm}

\noindent Note that, keeping the central spacecraft intact, we obtain six different sets of basis vectors simply by varying $x$ and $z$ directions (see Table \ref{table1}). Expressing all the calculated curls in $\left \{\hat{i}, \hat{j}, \hat{k}\right \}$ through the inverse transformation matrix, the final value of each GSE component of the curl is obtained by averaging over all the six choices \footnote{Note that, only the choice of coordinates in the current octant of the tetrahedron is considered to be consistent with the approximated expressions in Eqs. \eqref{eq:seta} -- \eqref{eq:setf}.}. The current density $\bm{J}$ is thus obtained using  $\bm{J} = \left ( \bm{\nabla} \times \bm{B} \right ) / \mu_0 $ whereas using the ion fluid velocity $\bm u_p$, one can calculate the ion vorticity as $\bm \omega_p = \bm {\nabla \times u}_p$.
{\renewcommand{\arraystretch}{1.4}
\begin{table}[!hbt]
        \centering
        \caption{Possible orientations for basis vectors keeping MMS 1 as the origin of the new coordinate system.}
        \begin{tabular}{ccc}
            \hline\hline
            Orientation & $\hat{x}$ axis & $x$-$y$ plane vectors \\
            \hline
            1 & $\hat{R}_{21}$ & $\hat{R}_{21}$ and $\hat{R}_{31}$ \\
            2 & $\hat{R}_{21}$ & $\hat{R}_{21}$ and $\hat{R}_{41}$ \\
            3 & $\hat{R}_{31}$ & $\hat{R}_{31}$ and $\hat{R}_{21}$ \\
            4 & $\hat{R}_{31}$ & $\hat{R}_{31}$ and $\hat{R}_{41}$ \\
            5 & $\hat{R}_{41}$ & $\hat{R}_{41}$ and $\hat{R}_{21}$ \\
            6 & $\hat{R}_{41}$ & $\hat{R}_{41}$ and $\hat{R}_{31}$ \\
            \hline
        \end{tabular}
        
        \label{table1}
    \end{table}}

In addition to curl, using IFM, it is also possible to calculate the divergence of a vector (say, $\bm A$) at MMS 1,
as,
\begin{align}
\bm{\nabla} \cdot \bm{A} {\big |}_1 & =
\dfrac{\partial A_x}{\partial x} + \dfrac{\partial A_y}{\partial y} + \dfrac{\partial A_z}{\partial z} \nonumber\\
& = \dfrac{{A_2}_x - {A_1}_x}{\mathbf{R}_{21} \cdot \hat{x}} +  \dfrac{{A_3}_y - {A_1}_y}{\mathbf{R}_{31} \cdot \hat{y}} + \dfrac{{A_4}_z - {A_1}_z}{\mathbf{R}_{41} \cdot \hat{z}}. \label{eqn8}
\end{align}
\noindent The calculated divergences, being scalar quantities, are averaged over the six choices to obtain the final value, without transforming them to GSE coordinates.
In order to apply IFM to the three-spacecraft system, we suppress one of the four spacecraft and are left with only one triangular surface. Keeping one of the vertices of the triangle as the centre, the $x$-axis can be chosen to be along any of the two sides common to the centre. The normal to this plane becomes the $z$-axis and $y$-axis is subsequently obtained to define the right-handed coordinate system. For example, let us suppress MMS 4 and work with the plane $\Delta 123$. Since all the spacecraft are in $x$-$y$ plane, no variation along $z$-axis is considered, thus leading to $\partial/\partial z \equiv 0$ and one can effectively write
\begin{equation}
\bm{\nabla} \times \bm{A} {\big |}_1 =
\left(\dfrac{\partial A_z}{\partial y} \right) \hat{x} + \left( - \dfrac{\partial A_z}{\partial x} \right) \hat{y} + \left(\dfrac{\partial A_y}{\partial x} - \dfrac{\partial A_x}{\partial y} \right) \hat{z}. \label{eqn9}
\end{equation}
Unlike the four-spacecraft system, here we have only two possible sets of basis vectors (orientations 1 and 3 in Table \ref{table1}) over which the final averaging is done.

\section{Results} 
\label{sec4}
\subsection{Calculation of spatial derivatives}
\begin{figure}
    \centering
    \includegraphics[width=0.95\linewidth]{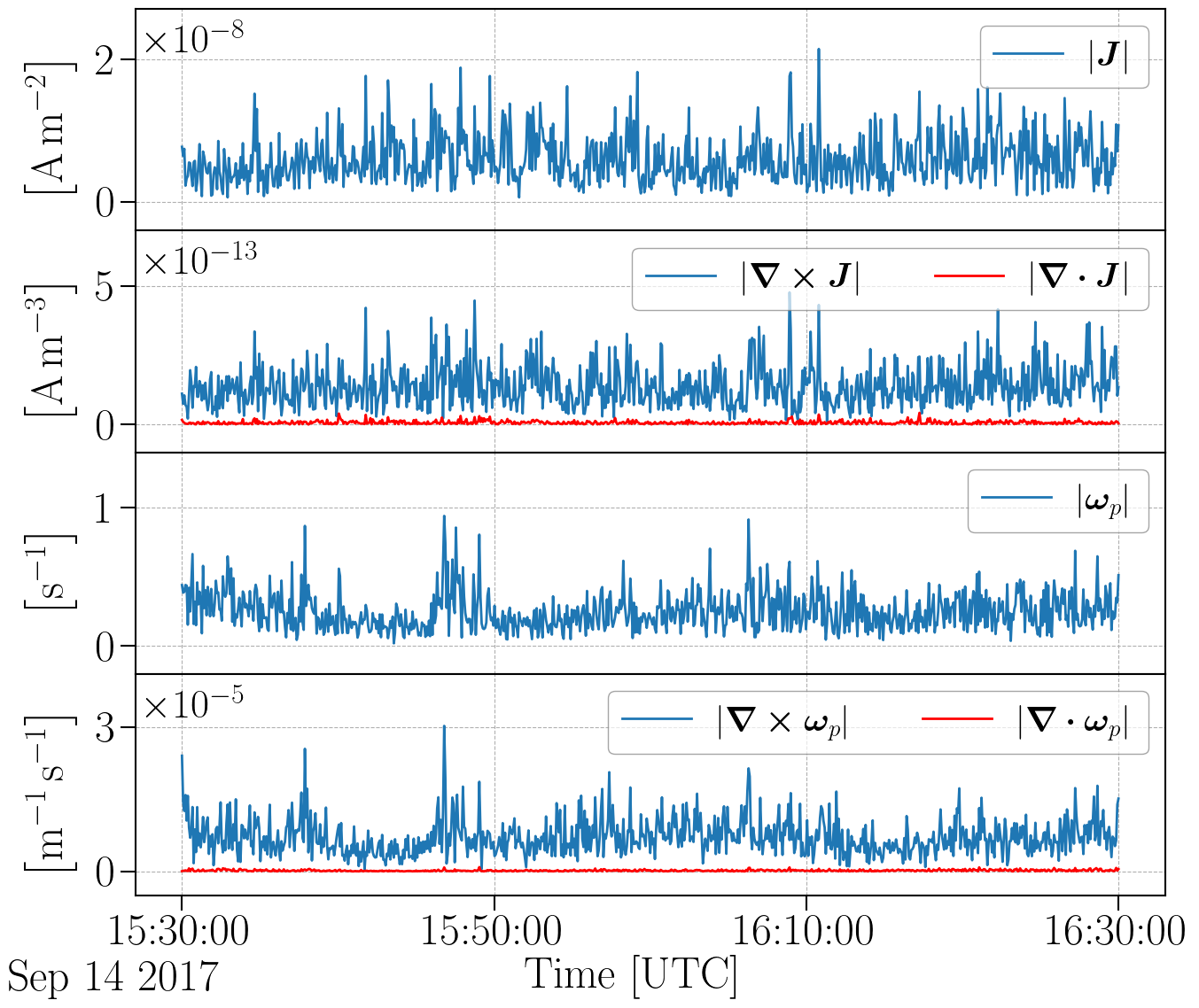}
    \caption{The magnitudes of $\bm{J}$ and $\bm{\omega}_p$ along with their curls and divergences calculated using IFM at $\bm{R}_1$ for 1 hour interval on September 14, 2017. }
    \label{jwplots}
\end{figure}

For the interval on September 14, 2017 from 15:30:00 UTC to 16:30:00 UTC, using IFM, we calculate $\bm J$ and $\bm {\omega}_p$ and plot their magnitudes (see Fig. \ref{jwplots}). As mentioned in Section \ref{sec1}, IFM allows the calculation of higher order derivatives of the curls. We therefore calculate the curls and divergences of the calculated curls and compare their magnitudes. The magnitude (GSE) of $\bm J$ is found to be $\sim 10^{-8} \text{A m}^{-2}$ whereas $\bm{\nabla \cdot J}$ has a magnitude $\sim 10^{-15} \text{A m}^{-3}$ which is 100 times smaller than $\left | \bm{\nabla \times J} \right | \sim 10^{-13} \text{A m}^{-3}$. The negligibly small value of $\left | \bm{\nabla \cdot J}\right |$, with respect to $\left | \bm{\nabla \times J}\right |$, is in accordance with the fact that $\bm J$ is the curl of a vector. For $\bm \omega_p$, the magnitude is found to be $\sim 10^{-1} \text{s}^{-1}$ whereas $\bm{\nabla \cdot \omega}_p$ has a magnitude $\sim 10^{-7} \text {m}^{-1}\,\text{s}^{-1}$ which is again 100 times smaller than $\left | \bm{\nabla \times \omega}_p\right |\sim 10^{-5} \text {m}^{-1}\,\text{s}^{-1}$, thereby supporting the fact that the divergence of a curl is zero. The other spacecraft locations, where we similarly calculate $\bm J$ and $\bm \omega_p$ along with their divergences and curls, also show the same behaviour.

\subsection{Application to three-spacecraft system}
\begin{figure}
    \centering
    \includegraphics[width=0.9\linewidth]{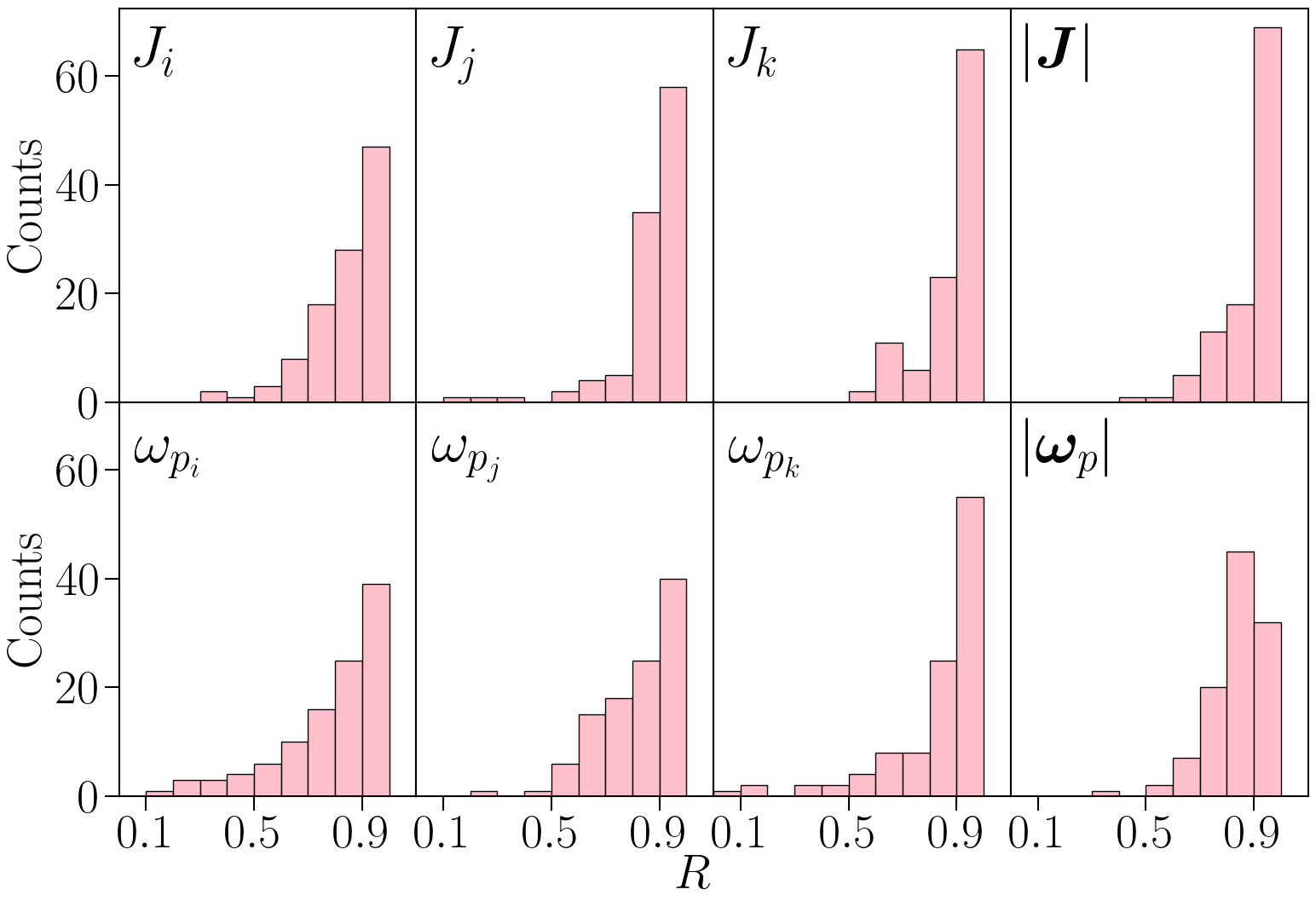}
    \caption{Histograms of correlation coefficients between quantities calculated using IFM (3 s/c) and IFM (4 s/c) at $\bm{R}_1$ for 107 datasets. The three-spacecraft system is obtained by suppressing MMS 4.}
    \label{hist3sc}
\end{figure}

To examine the applicability of IFM to a three-spacecraft system, we calculate $\bm J$ and $\bm \omega_p$ at the location of MMS 1 upon suppressing one of the remaining spacecraft (here, MMS 4) for a total of 107 intervals. To compare the curls calculated with and without suppressing MMS 4, we plot the histograms of correlation coefficients ($R$) between them for each component and the magnitude (see Fig. \ref{hist3sc}). Each histogram peaks at $R \sim 0.9$ indicating a reasonable agreement between the two calculations. In Table \ref{t3}, we enlist the mean correlation coefficients $\left \langle R \right \rangle $ and the corresponding standard deviations $\sigma_R$ for the three components and the magnitudes of $\bm J$ and $\bm \omega_p$. Despite a non-negligible spread $\sim 0.1$, which can be associated to the less number of possible coordinate systems, a reasonably high $\langle R\rangle \sim 80 \%$ justifies IFM to be a faithful way of curl calculation, even for a three-spacecraft system. This is an important advantage of IFM which can be very useful in a situation where one of the spacecraft data is unavailable or inappropriate for using.
\setlength{\tabcolsep}{0.5em}
{\renewcommand{\arraystretch}{1.3}
\begin{table}[!hbt]
    \centering
    \caption{${\left \langle R \right \rangle}$ between IFM (3 s/c) and IFM (4 s/c) with the corresponding standard deviations $\left (\sigma_R\right )$ for each component and magnitude of $\bm J$ and $\bm \omega_p$ at MMS 1, suppressing MMS 4.}
    \begin{tabular}{ccccccccc}
    \hline\hline
           &  $J_i$ &   $J_j$  &  $J_k$ & $\left |\bm{J}\right |$ &   $\omega_{p_i}$  &  $\omega_{p_j}$  & $\omega_{p_k}$ & $\left |\bm{\omega}_p \right |$\\
       \hline
        ${\left \langle R \right \rangle}$& 0.84 & 0.88 & 0.90 & 0.89 & 0.78 & 0.81 & 0.84 & 0.84 \\
        $\sigma_R$& 0.16 & 0.16 & 0.13 & 0.11 & 0.25 & 0.22 & 0.23 & 0.12 \\
    
        \hline
   \end{tabular}

   \label{t3}
\end{table}}

\subsection{Comparison with \textit{Curlometer} data}
\begin{figure*}
    \centering
    \includegraphics[width=0.9\linewidth]{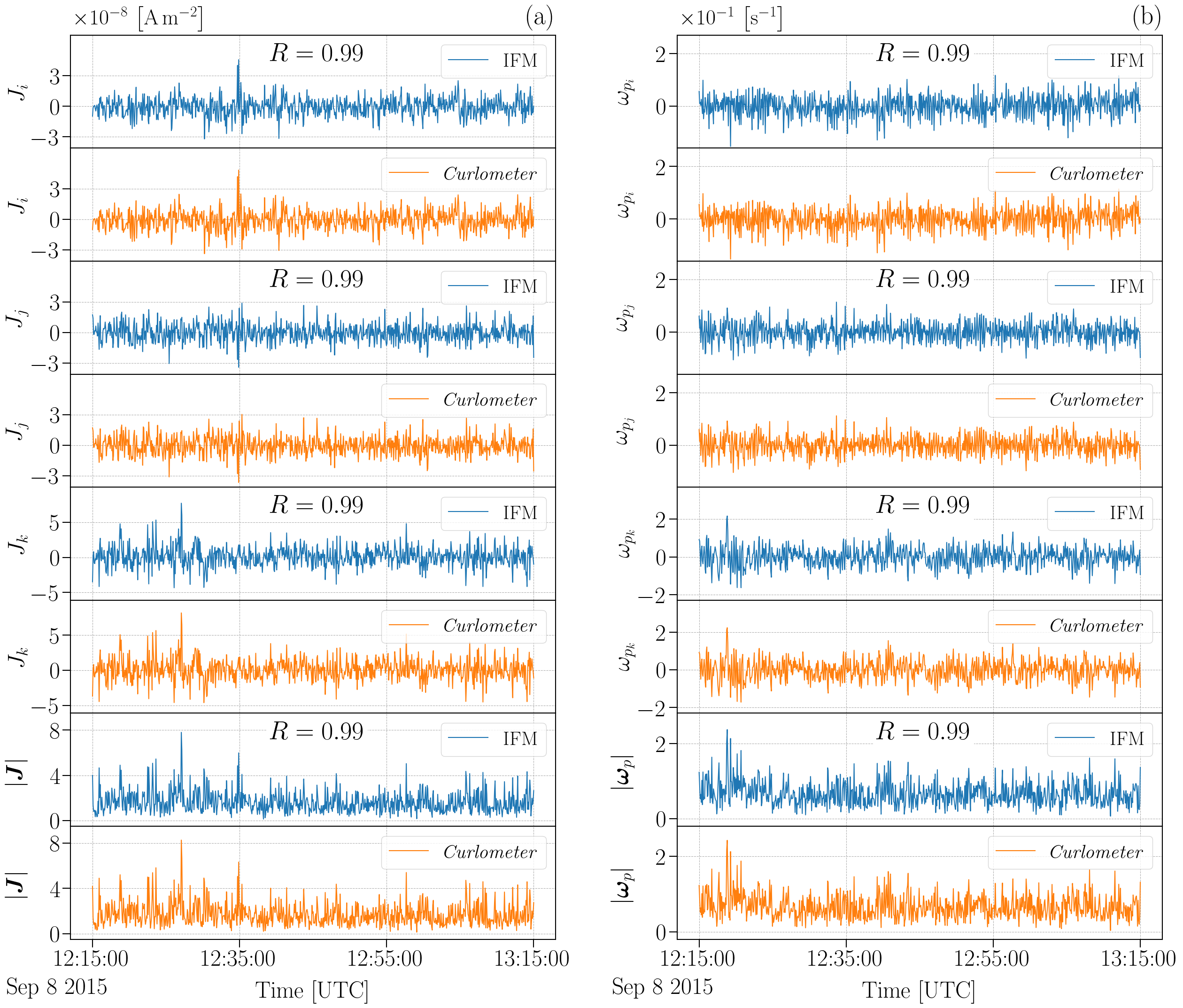}
    \caption{Comparison between averaged-IFM and \textit{Curlometer} calculation of (a) $\bm{J}$ and (b) $\bm{\omega}_p$ in terms of correlation coefficient $R$ for 1 hour interval on September 8, 2015.}
    \label{R}
\end{figure*}

In the next step, we compare our results with the curls calculated using the \textit{Curlometer} method. Note that, unlike IFM, \textit{Curlometer} provides an average value for the curl over the whole tetrahedron which does not correspond to the position of any specific spacecraft. The curl calculated using \textit{Curlometer} can therefore be compared with the average of those calculated over all the four spacecraft locations using IFM. A comparison of different components and the magnitude of $\bm J$ and $\bm \omega_p$ computed by these two methods is presented for September 8, 2015 [12:15:00 UTC - 13:15:00 UTC] in Fig. \ref{R}, which show an excellent agreement ($R \sim 0.99$) between these two methods. 

\begin{figure}
    \centering
    \includegraphics[width=0.9\linewidth]{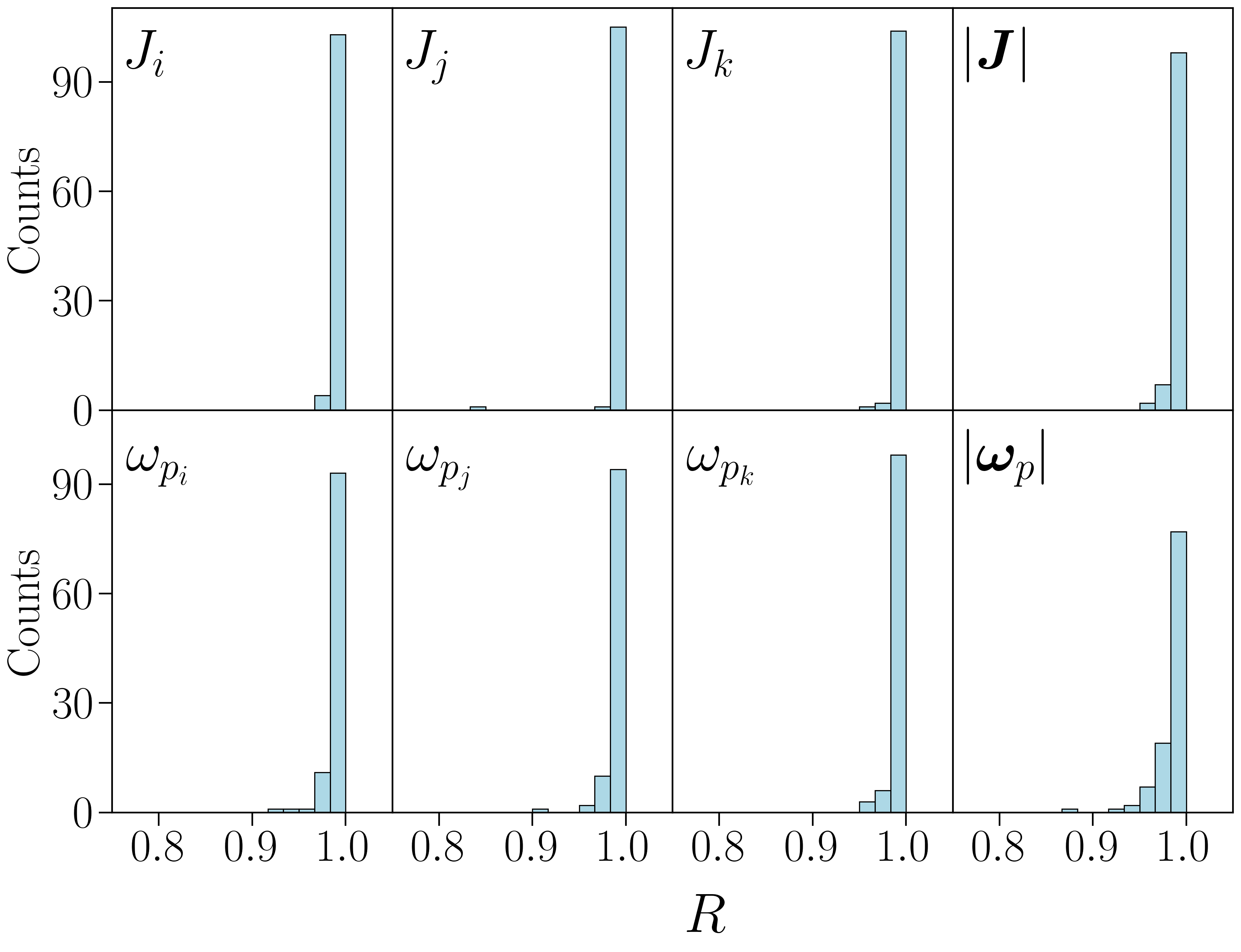}
    \caption{Histograms of correlation coefficients between quantities calculated using averaged-IFM and \textit{Curlometer} method for 107 datasets.}
    \label{hist}
\end{figure}
For all 107 datasets, we perform the same comparison between IFM and \textit{Curlometer} in terms of correlation coefficients ($R$). In Fig. \ref{hist}, we present the histograms for the correlation coefficients of each component of the curls and their magnitudes. For each case, the histograms are clearly peaking at $R = 0.99$ with a very narrow spread ($\sim 0.01$), thus showing an almost perfect match between \textit{Curlometer} and the averaged-IFM values. The mean correlation coefficients $\langle R \rangle $ and the corresponding standard deviations $\left (\sigma_R\right )$ for all the components of $\bm J$ and $\bm \omega_p$ are listed in Table \ref{t2}.
\setlength{\tabcolsep}{0.5em}
{\renewcommand{\arraystretch}{1.3}
\begin{table}[!hbt]
    \centering
    \caption{${\left \langle R \right \rangle}$ between averaged-IFM and \textit{Curlometer} with the corresponding standard deviations $\left (\sigma_R\right )$ for each component and magnitude of $\bm J$ and $\bm \omega_p$.}
    \begin{tabular}{ccccccccc}
    \hline\hline
           &  $J_i$ &   $J_j$  &  $J_k$ & $\left |\bm{J}\right |$ &   $\omega_{p_i}$  &  $\omega_{p_j}$  & $\omega_{p_k}$ & $\left |\bm{\omega}_p\right |$\\
       \hline
        ${\left \langle R \right \rangle}$& 0.99 & 0.99 & 0.99 & 0.99 & 0.99 & 0.99 & 0.99 & 0.98 \\
        $\sigma_R$ & 0.01 & 0.01 & 0.01 & 0.01 & 0.01 & 0.01 & 0.01 & 0.02
        \\
        \hline
   \end{tabular}
    
   \label{t2}
\end{table}}

\subsection{Determination of the quality of our method} \label{sec4_4}
\begin{figure}
    \centering
    \includegraphics[width= 0.85\linewidth]{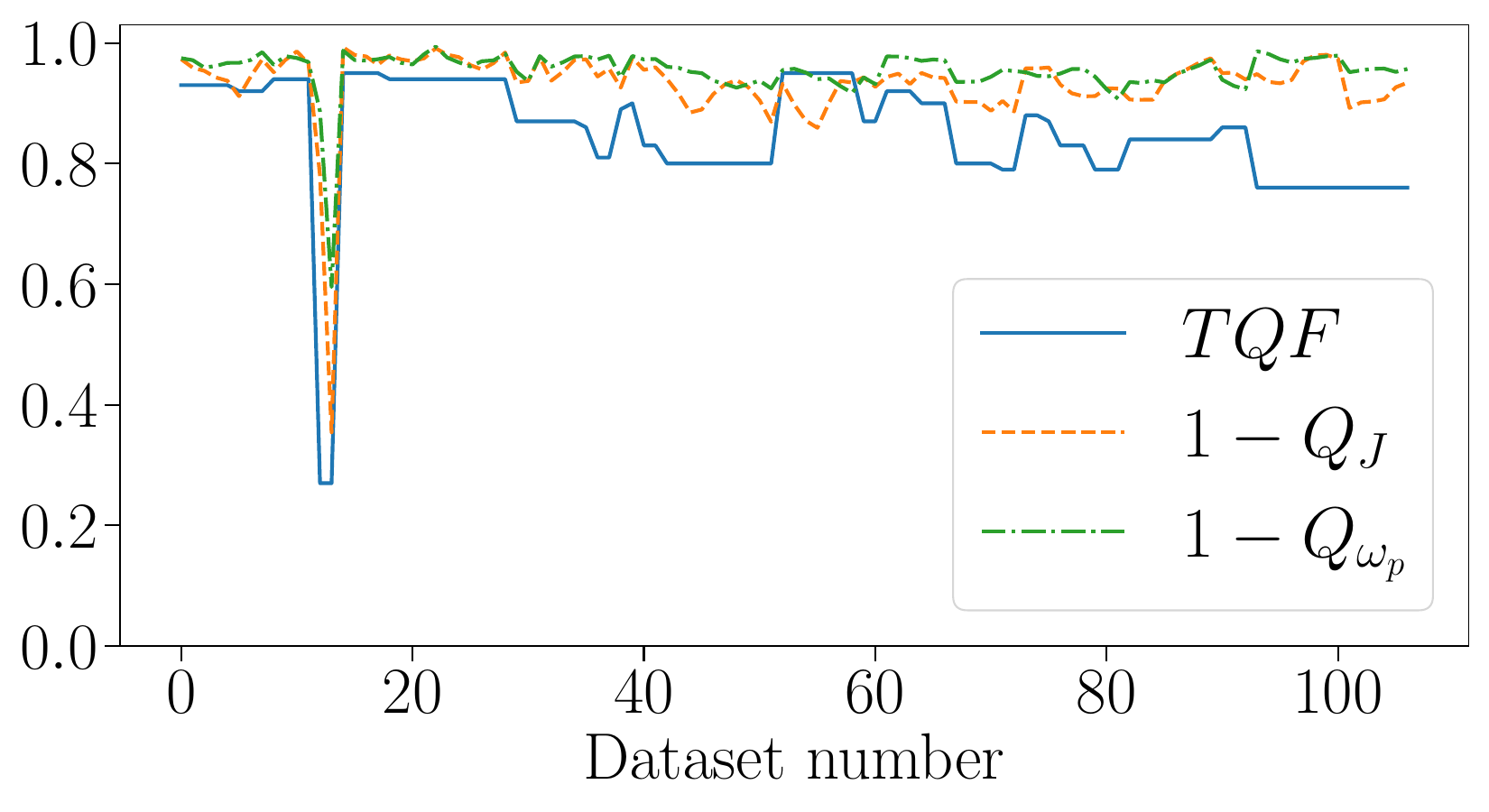}
    \caption{Variation of $1-Q_J$ and $1-Q_{\omega_p}$ with $TQF$ over all the datasets.}
    \label{tqf_qj_qw}
\end{figure}

The divergence of the curl of a vector is theoretically zero whereas the curl of the same need not be. Therefore, a possible measure of accuracy of the calculated curl ($\bm M$) may be quantified in terms of
\begin{equation}
    Q_M = \frac{\langle |\bm{\nabla \cdot M}| \rangle}{\langle| \bm{\nabla \times M}|\rangle} \label{qval},
\end{equation}
where the denominator is assumed to be non-negligible and here the average is taken over the entire time series during a given interval \citep{Dunlop_1988}. A small value of $Q_M$ corresponds to a better quality of curl and vice-versa. For all the intervals, we plot $1-Q_J$ and $1-Q_{\omega_p}$ in Fig. \ref{tqf_qj_qw} and show that both of them are considerably greater than 0.8, providing a good reliability of the calculated curl using IFM. The quality of the calculated curls is also dependent on the regularity of the tetrahedron (usually irregular for MMS), quantified in terms of the Tetrahedron Quality Factor, $TQF = Q_V Q_S$, where $Q_V$ is the quality factor for the tetrahedron shape defined as $Q_V = V_a/ V_r$ with $V_a$ being the actual volume of the tetrahedron and $V_r$ being the volume of a regular tetrahedron of side $L$ (equal to the average of the sides of the actual tetrahedron), and $Q_S$ is the quality factor for the spacecraft separation (refer Eqs. (3), (4) and (5) in \cite{Fuselier_2016} for definition). For a regular tetrahedron $TQF = 1$ and it becomes zero when the four spacecraft lie in the same plane. Fig. \ref{tqf_qj_qw} overplots $1-Q_J$, $1-Q_{\omega_p}$ and $TQF$ as a function of dataset number. In most of the cases, $1-Q_J$ and $1-Q_{\omega_p}$ are found to follow the trend of $TQF$. The correlation coefficients between $1-Q_J$ and $TQF$ is found to be 0.71 whereas that between $1-Q_{\omega_p}$ and $TQF$ is found to be $0.64$, showing a clear connection between the quality of the curl and the shape of the tetrahedron. In the same way, a sharp dip in $TQF$ between the datasets 10 and 20 is found to correspond to a sharp decrease in the quality of $\bm J$ and $\bm \omega_p$.

\section{Conclusion} \label{sec5}

In the current study, we propose an integration-free method for calculating curls and divergences of relevant vector fields (velocity, magnetic field, etc.) of space plasmas using the \textit{in-situ} data from multi-spacecraft missions (Cluster and MMS). For the four-spacecraft system, our method successfully calculates the derivatives (first order and above) at the location of each spacecraft for a total of 107 datasets. This is a significant advantage over the existing method of \textit{Curlometer} where only a single value for the entire multi-spacecraft system can be calculated. We obtain small values for the ratios between the magnitudes of the divergence and the curls of the calculated curls, aligning with the theoretical fact that the divergence of a curl vanishes. Unlike \textit{Curlometer}, our method can be used to calculate the curl and divergence even with a non-collinear three spacecraft system. A comparison with four spacecraft data shows a reasonably high correlation between the two, thereby making the calculation of the derivatives possible even when one of the spacecraft data is unavailable or inappropriate. For example, after facing flight anomaly in 2018, MMS 4 is unable to provide electron velocity data (see AMDA database) where IFM can be directly used to calculate its curl just by using the data from the rest of the spacecraft. However, the accuracy of the calculated derivatives using three spacecraft data is comparatively lower due to the less number of available orientations. Following the same line of argument, our method can also be applied to calculate the spatial derivatives for spacecraft missions containing more than four spacecraft (future HelioSwarm (NASA) mission, for example). However, the choice of the coordinate systems should be made depending on the instantaneous configuration of the system.

For all 107 intervals, a systematic comparison is made between the curls calculated using \textit{Curlometer} method with those obtained by the averaged-IFM (over four spacecraft). A very high correlation coefficient, $R \approx 0.99$ shows an excellent agreement between them, thus validating our proposed method over the popular benchmark of \textit{Curlometer}. For each dataset, we quantify the accuracy of the curls in terms of $1 - Q_J$ and $1 - Q_{\omega_p}$ which are found to be close to 1 indicating a good quality of curl calculation. We also connect the quality of curl with the regularity of the tetrahedron shape and find a reasonably good dependence between the two. The dependence on the tetrahedron shape imparts a drawback to the proposed method as it is unable to provide with good quality curl calculation for the cases of highly irregular tetrahedra. Unlike \textit{Curlometer}, our method can calculate the local derivatives at specific spacecraft locations, thus facilitating the calculation of the turbulent heating rate that involves the values of velocity and magnetic fields at each spacecraft location. The knowledge of $|\bm J|$ can be utilized to directly identify the sites of current sheets, thus complementing the existing methods \citep{Li_2011, Osman_2014}.

The calculation of spatial gradients using IFM further enables us to estimate turbulent heating rates using exact scaling relations derived for compressible MHD turbulence \citep{Banerjee_2013, Banerjee_2018}. In addition, an examination of point-wise $\bm{u}_p - \bm{\omega}_p$ and $\bm{B} - \bm{J}$ alignments (Beltramization) can be carried out providing insights into turbulent relaxation process \citep{Mahajan_1998, Banerjee_2023}.

\begin{acknowledgements}
     The authors acknowledge Space Technology Cell-ISRO project (STC/PHY/2023664O) for financial support. S. B. acknowledges Dibyendu Chakrabarty for useful discussions. Data analysis was performed with the AMDA science analysis system provided by the Centre de Données de la Physique des Plasmas (CDPP) supported by CNRS, CNES, Observatoire de Paris and Université Paul Sabatier, Toulouse. The values of $TQF$ used in Section \ref{sec4_4} are obtained from the website of MMS Science Data Center. 
\end{acknowledgements}

%
\bibliographystyle{aa} 
\bibliography{main} 
%

\end{document}